# From phonons to domain walls, the central peak and "critical slowing down"

Ekhard K.H. Salje[1] and Annette Bussmann-Holder[2]

[1] Department of Earth Sciences, University of Cambridge, Cambridge CB2 3EQ, UK
[2]Max-Planck-Institute for Solid State Research, Heisenbergstr. 1, D-70569 Stuttgart, Germany

**Abstract**

We investigate perovskite oxides from different aspects, namely their pseudoharmonic dynamical properties, their dynamical properties when strong anharmonicity exists and the intriguing functionalities arising from domain walls. Taking these viewpoints together yields a rather complex picture of this material class which was not anticipated in previous approaches. It opens pathways to novel applications and reveals the rich ground states beyond the fictitious belief in 'simplicity of perovskites and such structures'.

**Introduction**

Perturbation approaches have dominated the field of phase transitions ever since the celebrated papers by Landau and Lifshitz since the late 1930s [1, 2]. When Bill Cochran in 1959 [3] first formulated the idea that parameters like temperature, pressure, chemical potential etc gently modify phonons, the concept of optical soft modes was born. In all displacive systems, phonons are weakly anharmonic, which leads to linear temperature dependences of the squared phonon frequency $\omega^2$. In the high symmetry phase, the linear $\omega^2$ dependence is virtually always observed over relatively large temperature intervals while phonon branches at lower temperatures split in the symmetry-broken phase. Additional anharmonicities become apparent in the symmetry broken phase. In second order phase transitions, as exemplified in $SrTiO_3$, an additional conundrum was already discussed by Alex Müller: if the phonon frequency approaches zero the phonon amplitude diverges. Terms like 'critical slowing down' for the motion and 'central peak' for the frequency spectrum were coined to describe this situation – often without identifying the exact physical process which stabilizes phonons in the close vicinity of the transition point. Here we describe these phenomena in simple models without recurring to DFT techniques [4] in order to emphasise the underlying physical principles in the most transparent way.

**Pseudo-harmonic approach**

The first theories devoted to the lattice dynamics in ferroelectric perovskites can be traced back to Cochran (1961) [5], where he introduces a shell model description for ferroelectrics to account for the electronic polarizability of the ions, i.e. they are not describable as rigid ions but the relative displacements of their electronic shells with respect to the core have to be included. By using the adiabatic principle, the electronic and ionic degrees of freedom are decoupled, and the renormalized ionic model be considered in deeper detail. In view of the discovery that in perovskite oxides the ferroelectric phase transition is accompanied by the softening of an optic mode, the model explicitly considers anharmonicity by introducing a fourth order term in the lattice dynamical Hamiltonian [6, 7]. The linear T dependence of the squared soft optic mode is obtained within the self-consistent phonon approximation (SPA) where the fourth order term is replaced by its cumulant expansion [8]. However, a closer look at the temperature evolution of $\omega^2$ reveals that such simple dependence is not realized, neither in the limit of $\omega^2 \to 0$ nor at $T \gg T_C$, where saturation of $\omega^2$ is observed [9]. In the limit $\omega^2 \to 0$ very apparent deviations from the above law are commonly seen in quantum paraelectrics like e.g. $SrTiO_3$ [10] where saturation of the soft mode takes place due to quantum fluctuations which are larger than the original soft mode which is related to the displacement coordinates. Both temperature regimes have been successfully described within the polarizability model [11] where an anisotropic core-shell coupling at the oxygen ion is explicitly incorporated [12, 13]. With respect to the rigid A ion in $ABO_3$ this coupling is purely harmonic. Regarding the B ion the coupling consists of an attractive harmonic term $g_2$ and a stabilizing anharmonic fourth order term $g_4$ which within the SPA is replaced by a temperature dependent pseudo-harmonic approximation, namely: $g_T = g_2 + 3g_4\langle w_T^2\rangle$ where $\langle w_T^2\rangle$ is the relative core-shell displacement coordinate at the oxygen ion lattice site. The Hamiltonian of the polarizability model is given by

$$H = T + V \qquad (1a)$$

$$T = \sum_n [m_1 \dot{u}_{1n}^2 + m_2 \dot{u}_{2n}^2 + m_{el} \dot{v}_{1n}^2] \qquad (1b)$$

$$V = \sum_n [f'(u_{1n} - u_{1n-1})^2 + f(u_{2n} - w_{1n} - u_{1n})^2 + f(u_{2n+1} - w_{1n} - u_{1n})^2 + g_2 w_{1n}^2 + g_4 w_{1n}^4] \qquad (1c)$$

where $u_{in}, v_{1n}$ are the displacement coordinates of ion $m_i$ (i=1, 2) and shell $m_1$ in the n-th unit cell and $w_{1n} = v_{1n} - u_{1n}$, $f$ and $f'$ are nearest and second nearest neighbour harmonic coupling constants. Within this approach the lowest transverse optic and acoustic modes are well described as a function of momentum q and temperature T, where the latter is obtained through the SPA. The soft transverse q=0 optic mode is explicitly given by: $\mu \omega_{TO}^2 (q = 0) = 2 f g_T / (2f + g_T)$, with $\mu$ being the reduced cell mass. While the conventional approach attains the linear in T behaviour [9] for any temperature, this equation reproduces the quantum paraelectric behaviour as well as the saturation limit due to the T-dependence of the denominator. [Explicit consequences for quantum extensions to phase transitions were discussed in [2]] An interesting feature appears in the limit when optic and acoustic modes start to couple, which happens as function of temperature and momentum [14, 15, 16]. In such a case a critical momentum $q_c$ can be defined where both modes are closest to each other. Its value moves to q=0 with decreasing temperature and defines real space regions which are identified with elastic anomalies and polar nanoregions. Both grow in size when approaching the phase transition temperature $T_C$ to finally coalesce to a true global polar state. The onset temperature for their formation does not coincide with a Burn's temperature $T_B$ [17], which is notoriously difficult to determine experimentally, but occurs well below estimated values of $T_B$. A precursor temperature is defined where the crossover between the high temperature saturation and the pre-transitional softening occurs. This temperature was predicted as T*≈1.1$T_C$. [18]. Other values, which are in the same order magnitude, stem from Molecular Dynamics simulations [19] and been discussed in MSB (between 1.01 $T_c$ and 1.4 $T_c$). These temperatures have been observed experimentally in many perovskite oxides by Brillouin scattering [20], dielectric spectroscopy [21], birefringence [22], ultrasound, RUS of elastic softening [23 – 26] and various other techniques. Importantly, it demonstrates that a nominally displacive system always carries an order-disorder component and thereby indicates a coexistence of both classification classes [18, 27]. As is evident from equ. 1c a local double-well potential is inherently present in perovskite oxides which stems from the oxygen ion nonlinear polarizability which diverges as a function of its volume and is temperature dependent [28, 29].

**Periodons and Incommensurations**

In the following, the concept of soft modes is extended to nonlinear solutions where the continuum aspect of the $\Phi_4$ model [30] is complemented by discrete solutions appearing on the lattice [31]. First exact nonlinear solutions are compared to SPA solutions for the case of ferroelectric SnTe. [32, 33]. Next the coupling between the nonlinear and SPA phonons is discussed which leads to phase transitions at finite critical momentum $q_c \neq 0$ which is applicable to $K_2SeO_4$. In analogy to perovskites ionic systems are in the focus which can be modelled within the dipolar shell model approximation (see above). The harmonic lattice potential is given by: $\varphi^{(2)} = \varphi_{ii} + \varphi_{ei} + \varphi_{ee\prime}$ where $\varphi_{ii}$ is the ion-ion, $\varphi_{ei}$ the electron-ion and $\varphi_{ee\prime}$ the electron-electron interaction. In analogy to the case discussed above, the electron-ion interaction potential is extended by a fourth order anisotropic term $\varphi_{ei}^{(4)} = \frac{1}{4}\sum_{L,\alpha} g_{4,\alpha}(\kappa) w_\alpha^4(L)$ where $w$ is the relative core-shell displacement coordinate at lattice site $L = (l, \kappa)$ and $\alpha = x, y, z$. Within the SPA this term is replaced by: $w_\alpha^3(L) = 3w_\alpha(L)\langle w_\alpha^2(L)\rangle_T$ with the bracketed term being the thermal average. While in the continuum limit kink-type and solitary solutions exist, here nonlinear periodic 3 dimensional lattice solutions (periodons) are obtained using the ansatz for the displacements : $x = u, v, w$:

$$\vec{x}(L) = Re\{\vec{X_1}exp[i(\omega t - \vec{q}\vec{R}(L)] + \vec{X_3}exp[3i(\omega t - \vec{q}\vec{R}(L)]\}, \qquad (2)$$

with the amplitudes $\vec{X_1}, \vec{X_3}$ being determined by the equations of motion. The periodon dispersion relation is given by $\omega_p(\vec{q}) = \frac{1}{3}\omega_R(3\vec{q})$ where $\omega_R(\vec{q})$ is the SPA dispersion relation. In Fig. 1 periodons and phonons are shown for the ferroelectric rock-salt structure IV-VI compound SnTe. By including the interaction between phonons and periodons finite momentum soft modes are obtained which freeze at momentum q=$q_c$. This is best exemplified for the case of a pseudo-linear model with nearest neighbour interactions and one polarizable ion in the unit cell. This model is a simplification of the polarizability model, but capture its essential features, namely the nonlinear polarizability of the oxygen ion. Correspondingly, its Hamiltonian is only slightly varied as compared to equ. 1:

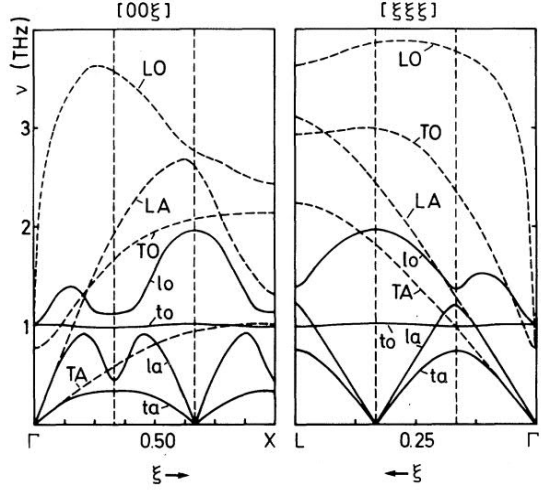

**Fig. 1** Dispersion curves of phonons (dashed lines) and periodons (solid lines) in SnTe at 100K. Capital (small) letters denote the polarization of them. Parameters have been adopted from Refs. [32, 33]

$$H = \frac{1}{2}\sum_n \left[ M_1 \dot{u}_{1n}^2 + m\dot{v}_{1n}^2 + f'(u_{1n} - u_{1n-1})^2 + f(v_{1n} - v_{1n-1})^2 + g_2 w_{1n}^2 + \frac{1}{2} g_4 w_{1n}^4 \right] \quad (3)$$

Where the same notations as above have been used. Since here only a single ion approximation is used, the index 1 is omitted in the following. The equations of motion are given by:

$$M\ddot{u}_n^2 = g_2 w_n + g_4 w_n^3 + f' D u_n \quad (4a)$$
$$m_e \ddot{v}_n = -g_2 w_n - g_4 w_n^3 + fD(w_n + u_n) = 0 \quad (4b)$$

Where $Dx_n = x_{n+1} + x_{n-1} - 2x_n$ is the difference operator. Solutions to these equations are given by: $w_n = A\sin(\omega t - nqa)$, $u_n = B\sin(\omega t - nqa) + C\sin 3(\omega t - nqa)$ with the dispersion relation: $M\omega_p^2(q) = \frac{4}{9}(f + f')\sin^2(\frac{3qa}{2})$ and amplitudes $A^2(q) = \frac{4}{3}\left[-g_2 - M\omega_f^2(1 - \frac{\omega_f^2}{\omega_R^2 - \omega_p^2})\right]$ with $\omega_f^2, \omega_R^2$ being the ferroelectric and rigid ion squared frequencies. The coupling between phonon and periodon is imposed by a superposition of their displacements, namely: $w_n = w_{np} + w_{ns}$, $u_n = u_{np} + u_{ns}$. This leads to two types of equations of motion, namely one in the periodon displacement coordinate $u_{np}$ and the other in the phonon related one $u_{ns}$:

$$M\ddot{u}_{ns} = (f + f')Du_{ns} + fDw_{ns} \tag{5a}$$
$$0 = [g_T + 3g_4 w_{np}^2] - fD[w_{ns} + u_{ns}] \tag{5b}$$
$$M\ddot{u}_{np} = (f + f')Du_{np} + fDw_{np} \tag{5c}$$
$$0 = g_T w_{np} + g_4 w_{np}^3 - fD[w_{np} + u_{np}] \tag{5d}$$

where for $w_{ns}$ the SPA has been used, and $g_T$ has been defined above. As is apparent from the above equations, the periodon amplitude adopts a temperature dependence where two regimes can be differentiated, namely the high temperature one where SPA phonons exist in a fluctuating periodon field, whereas at low temperature static periodons are observed. This yields a site dependent electron-ion coupling:

$$g_T + 3g_4 w_{np}^2 \left(\frac{2\pi}{3}\right) = \begin{cases} g_T, n \equiv 0 \ (mod\,3) \\ -2g_T - \frac{9}{\frac{1}{f}+\frac{1}{f'}}, n \equiv 1 \ or \ 2 \ (mod\,3) \end{cases} \tag{6}$$

In Fig. 2, 3 the theoretical results for K$_2$SeO$_4$ are compared to experimental data [34, 35]. The temperature dependence is given by $g_T$ and concerns the commensurate part, whereas $f'(T)$ determines the incommensurate intersite elastic coupling which shifts the minimum in the dispersion $\omega(q)$ away from the commensurate value $qa = 2\pi/3$ to higher values.

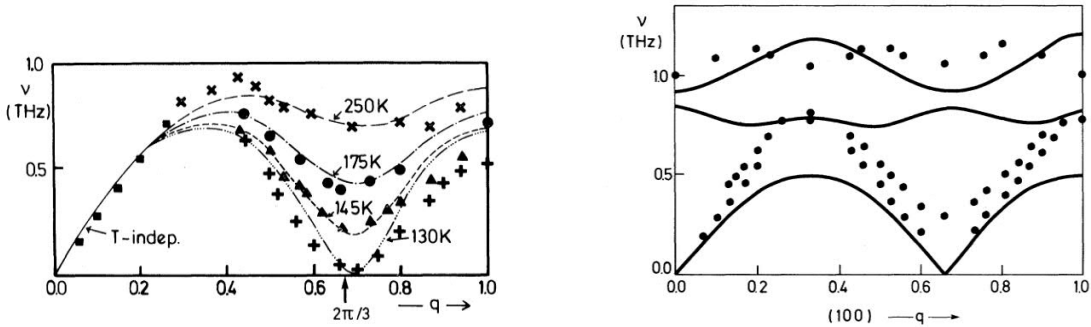

**Fig. 2** (left) Dispersion of the coupled phonon-periodon mode in the paraelectric regime (T>T$_C$=127K) of K$_2$SeO$_4$ with f+f'=09.THz$^2$x mass unit. Experimental data (crosses, dots, triangles, and crosses) have been taken from Refs. 34, 35.

**Fig. 3** (right) Tripling of the transverse acoustic coupled mode in the ferroelectric regime at T=40K.

Upon extrapolating $g_T$ and $f'(T)$ to zero (Fig. 4), it is obvious that the former governs the second order phase transition at $T_i$=127K, while the latter determines the lock-in transition at $T_C$=93K. In view of the simplicity of the model, the agreement between experiment and theory is amazingly good.

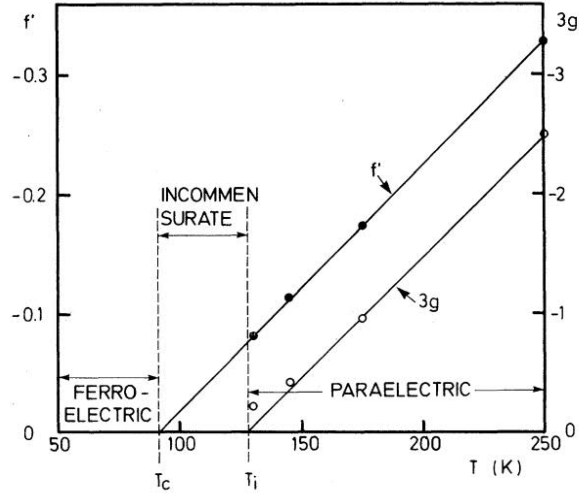

**Fig. 4** Temperature dependence of the coupling parameters $g_T$ and $f'(T)$ in the paraelectric and incommensurate regimes.

**Breather solutions and relaxor ferroelectrics**

Since many years it is well-known that perovskite oxides exhibit an enormous variety of ground states, induced by oxygen doping or deficiencies and/or by the replacements of the B or A sites by isovalent ions. The doped and mixed compounds are a platform for applications and new technologies. For these reasons increased scientific research has focussed on their local and global properties where also novel tools have been employed. Especially, it has been shown that their average global and local structures vary considerably [36 – 40]. This implies that the Bloch theorem is no more fulfilled and conventional approaches can no more be used. This essentially intrinsic heterogeneity infers that large anharmonicity is inherent to these compounds which suggests the formation of new intrinsic local modes (ILM) and discrete breather (DB) bound states [41 – 44]. The fascinating properties of these nonlinear solutions are that they are sources of energy localization and transport in nonlinear media.

A fundamental problem related to ILM and DB states is to confine them to an energetically existence regime, namely the gap region between optic and acoustic phonon modes. In addition, it has been argued that hard-core nonlinearity is not realizable physically [43]. The latter statement stems from the fact that a rigid ion or rigid spin-lattice potential has been considered for the calculations. Here we use instead the polarizability model as defined by equs. 1, where charge transfer, dynamical covalency and hybridization effects are explicitly included.

Instead of the above discussed periodon solutions the possible formation of DB is investigated [45] which requires that its frequency is constant within a limited spatial region and zero beyond that, which corresponds to multi-vibrational quanta bound state. In order to obtain the equations of motion the displacement coordinates are assumed to be time periodic:

$$u_{1n}(t) = A\xi_{1n}\cos(\omega t) \tag{7a}$$
$$u_{2n}(t) = B\xi_{2n}\cos(\omega t) \tag{7b}$$
$$w_{1n}(t) = C\eta_{1n}\cos(\omega t) \tag{7c}$$

with A, B, C being the amplitudes and $\xi, \eta$ the corresponding displacements. The amplitudes are obtained through the equations of motion and are frequency dependent, the electronic shells are treated in the adiabatic limit. The equations of motion are given by:

$$-m_1\omega^2\xi_{1n}A\cos(\omega t) = f'AD\xi_{1n}\cos(\omega t) + g_2C\eta_{1n}\cos(\omega t) + g_rC^{r-1}\eta_{1n}^{r-1}[\cos(\omega t)]^{r-1} \tag{8a}$$

$$-m_2\omega^2\xi_{2n}B\cos(\omega t) = fC(\eta_{1n+1} + \eta_{1n})\cos(\omega t) + fA(\xi_{1n+1} + \xi_{1n})\cos(\omega t) - 2fB\xi_{2n}\cos(\omega t) \tag{8b}$$

$$0 = -g_2C\eta_{1n} - g_rC^{r-1}\eta_{1n}^{r-1}[\cos(\omega t)]^{r-1} - 2fC\eta_{1n}\cos(\omega t) - 2fA\xi_{1n}\cos(\omega t) + fB(\xi_{2n-1} + \xi_{2n})\cos(\omega t) \tag{8c}$$

Here, r is the degree of anharmonicity. In order to reduce the number of degrees of freedom, equ. 8b is replaced by one lattice constant:

$$-m_2\omega^2\xi_{2n-1}B\cos(\omega t) = fC(\eta_{1n-1} + \eta_{1n})\cos(\omega t) + fA(\xi_{1n-1} + \xi_{1n})\cos(\omega t) - 2fB\xi_{2n-1}\cos(\omega t) \tag{9}$$

Adding eqs. 8b and 9 and twice the adiabatic approximation yields:

$$(m_2\omega^2 - 2f)\{g_2 C\eta_{1n} \cos(\omega t) + g_r C^{r-1}\eta_{1n}^{r-1}[\cos(\omega t)]^{r-1}\} + (m_2\omega^2 - 2f)[C\eta_{1n} + A\xi_{1n}]\cos(\omega t) = -f^2\cos(\omega t)[CD\eta_{1n} + AD\xi_{1n}] \quad (10)$$

Equs. 10 and 8a have analytical solutions which can be evaluated by the rotating wave approximation, namely:

$$\cos^p(\Theta) = \frac{p!}{2^{p-1}}\sum_{k=1}^{p} \frac{\cos(k\Theta)}{\left[\frac{p+k}{2}\right]!\left[\frac{p-k}{2}\right]!} \qquad p, k \equiv odd \quad (11)$$

$$\cos^{r-1}(\omega t) = C_r \cos(\omega t) + higher\ harmonics \quad (12)$$

with $C_r = \frac{(r-1)!}{2^{r-2}\left(\frac{r}{2}\right)!\left(\frac{r}{2}-1\right)!}$ where only the first term in the expansion is taken into account since all higher order terms rapidly decrease. The so-called worst case $r = 4$ is considered for systems with inversion symmetry, which excludes the case $r = 3$ and defines $C_r = 3/4$. The resulting potential is a hard core nonlinear one if $g_2 < 0, g_4 > 0$ and a soft one if $g_2 > 0, g_4 > 0$. With the harmonic coupling being attractive and the anharmonic coupling repulsive, hard core anharmonicity is given which might turn into a soft one when lattice and ILM solutions are superimposed.

The equations of motion are thus similar to those obtained from eqs. 1. Interesting novel solutions are obtained by the ansatz $w = -2u$ which yields complex results not discussed here. A rather natural choice for ILM's odd parity displacement patterns have been chosen, where at site $n = 0$ a dipole moment with "length" $n_c$ is formed which is compensated by the surrounding ordered lattice by creating at site $n$ counted from $n_c$ a dipole moment in the opposite direction with displacement $\xi_n$ according to $\xi_n = \xi_0 - \frac{n}{2}: n \equiv even;\ \xi_n = -\left[\xi_0 - \frac{n}{2}\right]: n \equiv odd$. This choice guarantees that at lattice site $n_c$ the breather extensions have reached their limit. The spread of the breather spatial extensions depends on the magnitude of the central dipole moment. With the choice $\eta_n = -2\xi_n$ two solutions for $\omega$ are obtained which need to be the same over all sites of the breather extent until $n_c$ is reached (Fig. 5 top). Figure 5 bottom provides an approximate description of the top figure in terms of a damped oscillator.

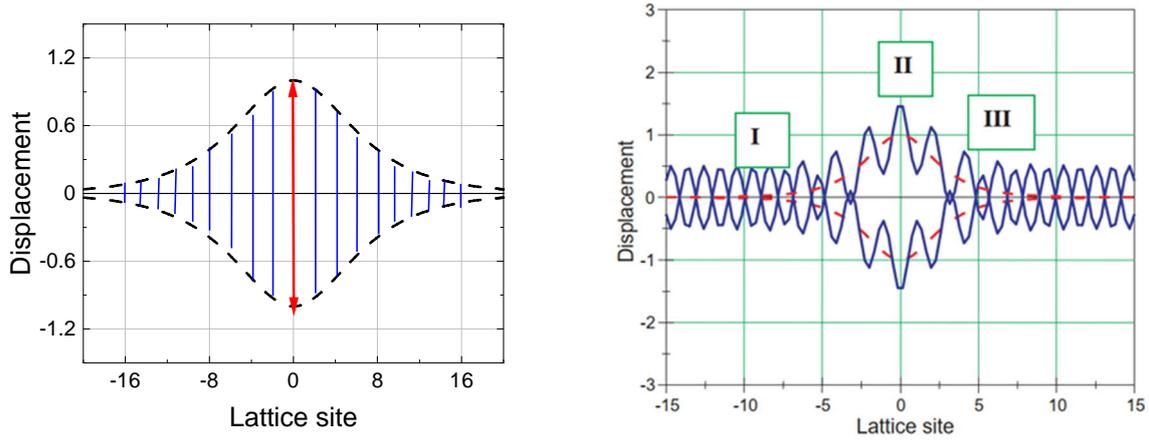

**Fig. 5** (left) Displacements of ion 1 with mass $m_1$ within the breather region. The breather center is marked by the red arrow. For clarity ion 2 is omitted. The black line corresponds to an envelope function proportional to 1/cosh and -1/cosh. (right) The displacement pattern stemming from the superposition of DB and regular lattice modes. I refers to the regular lattice modes, III defines a charge transfer region, II corresponds to the DB center. Region III is characterized by the mixing of DB and regular lattice displacement.

Thus, the breather spatial spread is $2n_c$ with small deviations for the n=0 solution when $g_4$ becomes site dependent like: $\tilde{g}_4^{(n)} = g_4/2(n-n_c)^2$. The result is consistent with the fact that the double-well potential is steep and broad in the breather center and becomes smoother with increasing $n$ to be pseudo harmonic for distances further apart. Simultaneously, this implies that in the centre the electron is far apart from its core, delocalized and spread over the whole DB extent, whereas at the boundaries it is completely localized. The scenario resembles closely polaron physics with the distinction that the electron is not trapped by the lattice distortion but moves almost freely through the distorted region.

The frequencies for the displacement patterns are given by:

$$\omega_1^2 = \frac{1}{m_1}\left[4f' + g\frac{C}{A}\right] - \frac{2f'}{n_c m_1} \tag{13a}$$

$$\omega_2^2 = \frac{2f}{m_2}\left[1 - \frac{1}{n_c}\frac{1}{2+\frac{gC}{2f(2C-A)}}\right] \tag{13b}$$

with $g = 2g_2 + C_r \tilde{g}_4 C^2$. Outside the DB region the frequencies are:

$\omega_1^2 = \frac{1}{m_1}\left(4f' + g\frac{C}{A}\right)$, and $\omega_2^2 = \frac{2f}{m_2}$. The existence regimes of the solutions equs. 13 are defined by comparing those to the SPA solutions. As already outlined above these need to lie in the gap region between optic and acoustic modes which excludes $\omega_2^2$ from further considerations. However, also $\omega_1^2$ is subject to certain restrictions since this lies only in the gap regions if the inequality $(2\tilde{f})^{1/2} < (g\frac{C}{A})^{1/2} < [\frac{2\tilde{f}(m_1+m_2)}{m_2} - 4f']^{1/2}$. Apparently, a decisive role for the DB formation is played by the ionic masses, elasticity related to f', and the lattice stability as determined by $2\tilde{f} = 2fg_T/(2f + g_T)$. Elastic softness clearly supports DB formation, whereas lattice stability does not contribute significantly. Analogous to the periodon solutions the interplay between the lattice and DB modes can be explored by a linear superposition of the corresponding displacement patterns. This leads to a stabilization of the phonon frequencies where even in the case of $g_T = 0$ the coupling to the breather prevents a freezing of any soft mode. This enlarges the DB existence regime. Conversely the breather adopts a temperature dependence since $g_2$ is replaced by $g_T$ which renormalizes its potential from hard-core to soft. This ansatz has the consequence that the dynamical interplay between the regular lattice and the locally distorted regions appear in a strongly enhanced diffuse scattering as, e.g., observed in relaxor ferroelectrics. Especially, ghost modes may be observed which take away spectral weight from the regular lattice modes at certain wave vectors as defined by the DB spatial extent. Three regions of the coupled modes can be differentiated (Fig. 5 right), I, the regular lattice modes, III a charge transfer regime where the lattice modes amplitude is modified by the breather, and, II, where the breather centre induces large distortions with almost freely moving charges. Since the nonlinearity, as defined by $g_4$, becomes site dependent, also the dielectric response $\varepsilon$ varies locally. This is what is observed in relaxor ferroelectrics, where $\varepsilon$ is frequency dependent, and indicates that ILM formation could be the origin [46, 47].

**Domains and domain walls**

An alternative approach was developed when the phonon anharmonicity is beyond perturbation theory and they transform into solitary waves [48,49]. These excitations are often better understood in a quasi-static context where their dynamics is considered separately from their phonon origins. The probably most obvious example for such solitary waves is a twin boundary. As an example, when $SrTiO_3$ undergoes the symmetry breaking phase transition near 105K the low temperature phase is commonly riddled with twin domains [50-53]. These twin domains were the bane of most research, and a common approach was to simply ignore them. This is justifiable because the twist between twin domains is small, and each twin domain follows very closely the same physics of any other twin domain or even of an entire, untwined sample. These 'bulk' properties are hardly influenced by the twins and twinning was considered a nuisance, at best. This is not true for the geometrical joints between the twin domains, namely the twin boundaries (or twin walls). Since 2010 research on twin walls has massively expanded because it was discovered that these twin walls house many new and useful physical properties, which the bulk does not possess. These 'emerging' properties include polarity, (ionic and super) conductivity, photovoltaic effect, and pn junctions [54-59]. There is another dynamic wall property which is widely used. Most twin walls are mobile under external forcing when pinning is weak [60-63]. This means that small external stresses, such as the tip of a preparation needle, will move domain walls to desired positions. Even without external forces, domain walls can be moved by thermal stresses depending on the shape of the sample [64-70]. Importantly, twin walls often interact. They also contain internal geometrical structures, namely kinks, which represent walls in walls and lead to additional excitations where such kinks move at supersonic speed [68]. We now focus on the dynamical features in the 'critical slowing down' regime in both phases and relate them to the interactions and correlations [49,61] between interfaces such as twin walls. The movement of correlated twin walls, with and without external forcing, often occurs in avalanches. Such avalanches are known to geophysicists from Earthquakes. Avalanches in nano-materials have come to the forefront and dominate much of the current research [71,72]. Their experimental incarnation is often called 'crackling noise' [73].

The concept of crackling noise has been around for a long time. It refers to the jerky response of many systems to a slowly changing driving force or field. For example, a piece of paper crackles when it is slowly crumpled and corn flakes crackle when you pour milk over them. Breaking a chop stick [74] or a piece of coal or sandstone leads to audible crackles [75] and can easily be measured by putting microphones next to the sample [76]. Other examples include the braking of bones [77] or the destruction of kidney stones in the operation theatre [78]. Earthquakes follow similar patterns [79- 81]. Similarly, magnetic materials magnetize via jumps in the magnetization that span a wide range of jump sizes.

These jumps were originally observed as crackling Barkhausen noise when a search coil was wrapped about the sample and hooked up to a loudspeaker [82, 83]. Starting with the Barkhausen analysis of magnetization jumps in slowly magnetized ferromagnetics, the concept of crackles with a broad (power-law) size distribution was generalized to crackling noise because it was found that similar phenomena are surprisingly widespread [71,73,85].

Anharmonic phonons transmute hence to domain walls which have, like phonons, some universal properties. Renormalization group calculations suggest that, on long length scales, the avalanche systems of domain walls flow to the same fixed point under coarse graining, which suggests that their scaling behaviour is the same for all systems [85, 86]. Important open questions concern the size of the underlying universality class, i.e., how many systems show the same crackling noise statistics. This is often quantified by power laws and scaling functions underlying the avalanche size and duration distributions, the power spectra of acoustic emissions (AEs), and related quantities. The power law distributions of crackling noise also imply that these processes are scale invariant: each measurement interval shows exactly the same functional form of the jerk probability. This scale invariance is absolutely stunning because in many large systems it extends over 8 decades of the crackling noise energy and is truncated only by the limitations of the electronic equipment which measures the different jerk parameters [e.g.71]. It is rare to find physical laws in solid state physics which have excellent validity over 8 energy decades! A large body of experimental studies have proven these ideas to be largely correct. The power laws of the various observables are very similar indeed, including the energy, amplitude and size exponents, and the duration of the excitation. The universality of the crackling noise is further focused because energy exponents in ferroic materials are often near the mean field (MF) values of 1.33 and 1.66 while higher values were measured more rarely in other systems [71].

In this chapter we bring forward that domain boundaries evolve out of anharmonic phonons. This gives the system an order-disorder component which coexists with the phonon degrees of freedom. Alternatively, starting from an order-disorder spin model, Salje and Dahmen [71] have argued that very simple theoretical models can describe domain wall avalanches rather well. Using magnets as an instructive case, the random field Ising model (RFIM), with or without added long-range dipolar interactions is a typical eaxample. In the RFIM, the material is modelled as a cubic crystal with N sites, each with a spin $S_i = +-1$, i= 1... N. The Hamiltonian for the RFIM is $H = -\sum_{ij} J_{ij} S_i S_j - (H_{ext}(t) +$ $h_i)S_i$, where the first term represents the ferromagnetically coupled nearest neighbours (with nearest neighbour coupling $J_{ij}$ [J > 0], $H_{ext}(t)$ is the external applied field, and $h_i$ is the quenched local random magnetic field with Gaussian distribution. In the simplest non-equilibrium dynamics, each spin is aligned with its local effective field, $h_{j,effective} == -\sum_{ij} J_{ij} S_i S_j + H_{ext}(t) + h_i)$. As the external magnetic field is slow, it triggers spins with positive random fields to flip from down ($S_j$=-1) to up ($S_j$ =+1) whenever their local effective field first becomes positive. Because of the ferromagnetic coupling between the spins, each such spin flip can trigger neighbouring spins to also flip, causing a spin-flip avalanche. In the adiabatic limit, the external driving field is kept fixed until an avalanche is completed. Only afterward is it increased until the next spin flips. We call S the size of a spin-flip avalanche. Two limiting cases are illustrative for avalanches. In a pure system with all random fields, one expects that all spins will flip in one giant spin-flip avalanche that covers the entire system. For infinite disorder, the random fields are so far apart that each spin flips separately, and there are no large spin-flip avalanches. The model predicts that between these two extremes at a critical disorder, a power-law avalanche size distribution [D(S)] is observed when the magnetic field is near a critical field ($H_c$). The power law is multiplied with an exponential cut off if H is tuned away from $H_c$ or the disorder is tuned away from the critical disorder. The predictions only depend on general properties, such as symmetries, dimensions, range of interactions, etc. Many more quantities have been predicted, but the scaling form of the avalanche size distribution is quite general. This picture appears valid for ferroic domain walls and has been advocated for the behaviour of $SrTiO_3$ [88] and $BaTiO_3$ [72].

For soft magnets, the coupling $J_{ij}$ is generalized to also include added long-range dipolar interactions and simplified power laws apply. Twin walls follow the same non-equilibrium pathways with typical avalanche exponents near mean field values. Changing the RFIM to a 'soft' spin model with long-ranging elastic interactions leads then to the same coexistence between phonon and domain walls.

**$SrTiO_3$ as an example**

$SrTiO_3$ undergoes a second order phase transition with one transverse acoustic soft mode at high temperatures which splits into 2 soft modes in the symmetry broken phase. In addition, domain walls appear, and their characteristics are shown in Fig. 6. Above the transition point near 106.5K a large precursor regime is seen by elastic softening (Fig. 6b). Below the transition point, domains form and increase the damping of acoustic wave (Fig. 6a). The domain walls freeze with a very small pinning energy and, on cooling, form clusters. These clusters give the crystal a glassy or solid damping behaviour with greatly diminished elastic Youngs modulus [88].

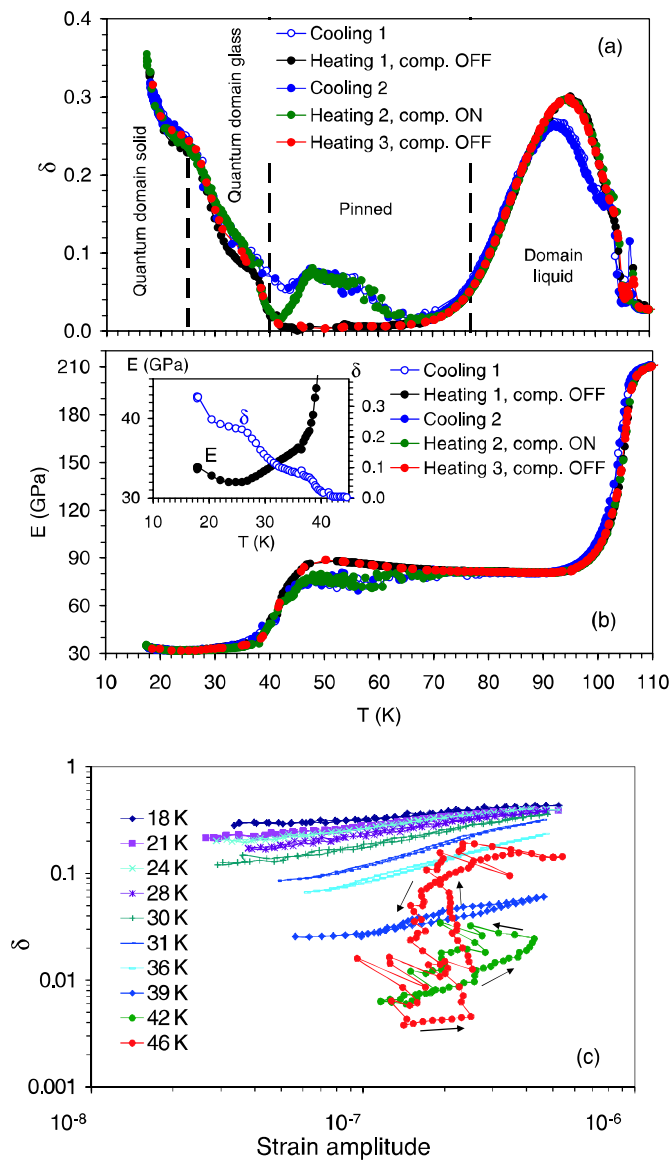

**Fig. 6** Domain wall characteristics in SrTiO$_3$ with (a) the damping of acoustic waves, (b) the elastic modulus, and (c) the damping as function of the applied strain. The strain amplitudes are extremely small and can be generated simply by heating or cooling samples with irregular shapes. Some unpinning already happens when the sample in a cryostat is cooled with the noise of the cryostat as sole unpinning force. Measurements have been made under 'silent' conditions [88]

In SrTiO$_3$, as expected, even more complex jerky avalanches were observed [88]. The movement of the twin walls under weak external forcing is shown in Fig.6. Close to the transition point, the walls are so weakly pinned that their mutual interaction dominates, and the domain wall pattern fluctuates like in a liquid. In the pinned regime, jerks and avalanches are observed and below 40K quantum effects destroy individual avalanches and lead to strongly coupled movements of clusters. The avalanches are seen as random jerks when the external strain exceeds some $10^{-7}$ and unpins the walls (Fig. 6c).

We now discuss the interaction between twin walls which is crucial for the avalanche formation [21]. The observed glassy behaviour [88] and the very existence of avalanches require such interactions, which appears counterintuitive following the work of Lajzerowicz and Levanyuk [89]. They and others showed that such interactions are extremely weak unless domain walls intersect [90]. While this argument is correct, we need to consider a further dimensional reduction. While domain walls are planes with some 1nm thickness, they host walls inside these wall [53]. These finer walls are often geometrically kinks which emit strong strain fields. These kink-kink strain fields interact like monopoles (~1/r where r is the distance between kinks) if the twin wall is inside the bulk and like dipoles (~1/r$^2$) close to surfaces. This weakening of interactions is due to elastic image forces and was shown experimentally by investigations of thin, freestanding sample lamellae [64, 68, 69, 91]. The long ranging strain interaction between kinks leads to effective wall-wall interactions like spin-spin interactions in spin glasses [92].

**Precursor Effects in SrTiO$_3$**

Anharmonic phonons become solitary waves and ultimately contribute to domain walls. Domain walls occur in the symmetry broken phase but very similar excitations are seen also in the high temperature phases. At temperatures above the transition point we find precursor clusters which have properties very similar to those of wall dominated systems in the symmetry broken phase. Experimentally, the measurement of the temperature evolution of the elastic moduli was very successful to demonstrate this precursor effect [93]. In a simple soft mode model the moduli transform stepwise in improper ferroelastics [94, 95] with no temperature evolution at T>$T_c$. This behaviour is hardly ever observed and typical lowering or enhancement of the moduli when approaching $T_c$ in a second order phase transition is commonly observed. The scaling of the elastic moduli follows a power law E ~ $E_0$ |T-$T_c$|$^{-\kappa}$ where κ is the precursor exponent. These exponents depend on the dispersion of the soft modes and hence on the dimensionality of the softening in reciprocal space. In a very simple model, values of κ have been predicted to be 1.5 if a single phonon branch flattens. If two orthogonal phonon branches flatten while the third remains relatively steep, we expect κ = 1. Finally, if three orthogonal branches flatten, the expected value is κ = 0.5. The experimental observations are surprisingly close to these values in $SrTiO_3$. The range of exponents is typically between 0.5 and 2 in most domain wall systems. The power-law dependence is well demonstrated with an exponent κ~ 0.2 for $BaTiO_3$ and 1.8 for $SrTiO_3$ [93]. Alternative Vogel-Fulcher fits

$$\Delta C_{ik} = B_{ik} exp\left(\frac{E_a/k_B}{T-T_{VF}}\right) \qquad (14)$$

are also shown in Fig.7. Here the activation energy is $E_a$ and the Vogel-Fulcher temperature is $T_{VF}$. Typically, the distinction between power laws and Vogel-Fulcher statistics is extremely difficult. In $SrTiO_3$, Cordero et al [93] have argued that the power law gives a better agreement with the experimental data. The precursor regime extends in $SrTiO_3$ to 125 K, other materials show similar of larger precursor temperature intervals [e.g. [96]].

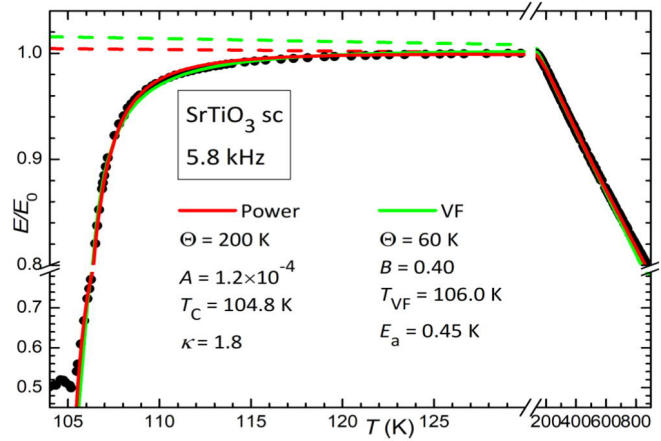

**Fig. 7.** Elastic precursor softening in SrTiO$_3$. The red and green lines represent the power law and the Vogel-Fulcher fits, respectively, the dotted lines are the baseline for the moduli at higher temperatures [after 93].

The softening of the elastic moduli in not purely a phonon effect. In a very simple model of an anharmonic Landau system [91], the crystalline structure inside the precursor regime was investigated by molecular dynamics simulations. The anharmonicity is a sheared NaCl type structure was simulated by anharmonic springs in the ball and strings model in Fig. 8.

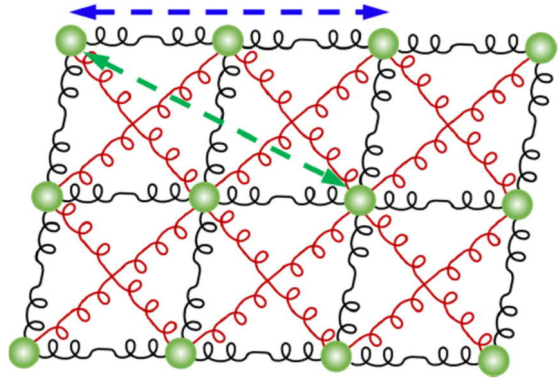

**Fig. 8**. Ball and spring model for phonons and domain walls. The red springs are Landau springs with two energy minima.

The potentials are

*Black springs* $U(r) = 0.1(r-1)^2$ (15a)

*Red springs* $U(r) = -0.05(r - 2)^2 + 40(r - 2)^4$ (15b)

along diagonals in the lattice unit, which has 2 energy minima and generates the Landau potential. The fourth-order third- nearest interactions:

$U(r) = 0.04(r - 2)^4$ (blue arrow), (15c)

and an anharmonic fourth-nearest Landau-type double- well interaction:

$U(r) = -0.05(r - \sqrt{5})^2 + 25.5(r - 5)^4$ (green arrow), (15d)

Models of this kind have been successfully used to investigate ferroelastic materials [97], phase transitions and reproduce both phonons and nano-structures very well [98]. They show that the correct softening as measured in $SrTiO_3$ and show the related nano-structure in Fig.9.

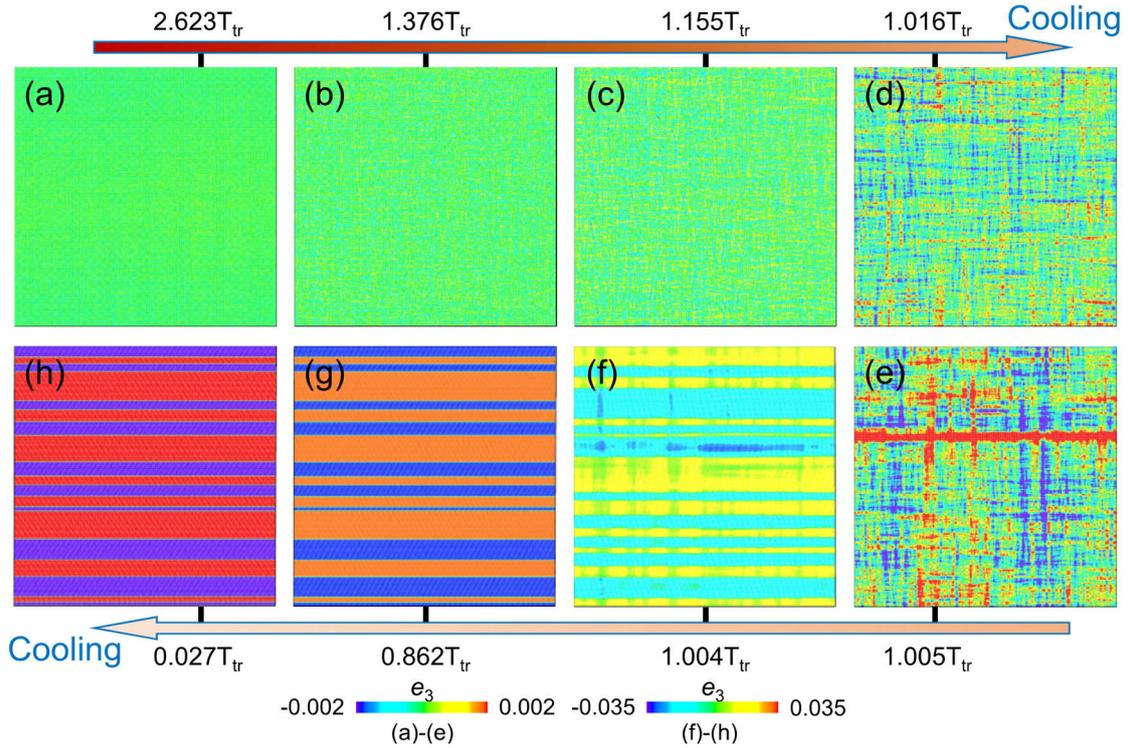

**Fig. 9**. Nano-structures, which spontaneously nucleate with the model in Fig.8. $T_{tr}$ is the transition temperature $T_c$ in $SrTiO_3$. In the symmetry broken phase, twin walls are clearly visible (g, h). At higher temperatures, wall-like nano-structures appear at (b) and become dominant in (d). These precursor structures reduce the relevant elastic modulus and link the anharmonic phonons at high temperatures with the domain walls at low temperatures [94].

This remarkable result shows that fine structures are visible at temperatures as high as 1.4 $T_c$ and become very strong at 1.016$T_c$ while the time averaged structure is firmly in the high temperature phase. The structural deformation resembles the typical tweed structure in order/disorder phase transitions [99] with a closely interwoven domain pattern. The domain structures seem to emerge in these structures although they are dynamic excitations which fluctuate and eliminate the actual symmetry breaking of the phase transition. Such excitations can be described as highly anharmonic 'phonon' branches or as solitary waves. When the actual phase transition occurs, stable domain walls appear (f-h) which are no longer related to phonon excitations.

**Summary and conclusions**

In summary, we have shown that the central peak and 'critical slowing down' regimes are characterised by the appearance of mobile, short-lived domain wall excitations which emerge out of phonon branches at higher temperatures. This idea refines the interpretation by Alex Müller where terms like 'entropic clusters' were used to describe what is rather like the features summarized in this paper. The emerging mechanism can then be depicted as follows: at T>>$T_c$ phonons are harmonic and become weakly anharmonic under cooling when the soft mode lowers the frequency. Above but near $T_c$ the phonon amplitudes increase and the frequencies collapse. This leads to highly anharmonic phonons which can be described by higher order dispersion terms like in periodons. Structurally, local clusters appear already at 1.4 $T_c$ in our simulations. They are widely separated. They grow and form denser patterns when approaching $T_c$. Characteristic temperatures can be defined to describe this process. While the lifetime of these clusters is much longer than the phonon time, they remain dynamic. At the transitions point, the clusters form domains with long-lived domain boundaries which contain still features of the precursor clusters. Lowering the temperature even further can lead (e.g. in $SrTiO_3$) to strong interactions between domains, domain walls and kinks inside domain walls, and to the formation of quantum domain glasses and quantum domain solids.